\documentclass[12pt,preprint,amsmath,amssymb]{article}
\usepackage{amssymb}
\usepackage{amsmath,amssymb,graphicx}
\usepackage{axodraw}
\usepackage{epsfig}
\usepackage{citesort}

\def\be{\begin{equation}}
\def\ee{\end{equation}}
\def\bea{\begin{eqnarray}}
\def\eea{\end{eqnarray}}
\def\lsim{\raise0.3ex\hbox{$\;<$\kern-0.75em\raise-1.1ex\hbox{$\sim\;$}}}
\def\gsim{\raise0.3ex\hbox{$\;>$\kern-0.75em\raise-1.1ex\hbox{$\sim\;$}}}
\def\ie{{\it i.e.}}
\begin{document}



\thispagestyle{empty}
\begin{center}
{\Large \bf Electron EDM and soft leptogenesis in supersymmetric
$B-L$ extension of the standard model\\}
\vspace{1.5cm} %

{\bf  Yuji Kajiyama$^1$, Shaaban Khalil$^{2,3}$ and Martti Raidal$^1$\\} %
\vspace{1cm} %


$^1$ \emph{National Institute of Chemical Physics and Biophysics,
Ravala 10, Tallinn 10143, Estonia.}\\

\vspace{0.3cm}

$^2$ \emph{Center for Theoretical Physics at the
British University in Egypt, Sherouk City, Cairo 11837, Egypt.}\\

\vspace{0.3cm}

$^3$ \emph{Department of Mathematics, Ain Shams University,
Faculty of Science, Cairo, 11566, Egypt.}\vspace{2.0cm}
\end{center}

\abstract{ 
We analyze the connection between electric dipole moment of the electron and the
soft leptogenesis in supersymmetric $B-L$ extension of the standard model. In
this model, the $B-L$ symmetry is radiatively broken at TeV scale. Therefore,
it is a natural framework for low scale seesaw mechanism and also for
implementing the soft leptogenesis. We show that the phases of trilinear
soft SUSY breaking couplings $A$, which are relevant for the lepton asymmetry, are not 
constrained by the present experimental bounds on electric dipole moment. 
As in the MSSM extended with right-handed neutrinos, 
successful leptogenesis 
requires small bilinear coupling $B$, which is now given by 
$A_N$ and $B-L$ breaking VEVs. 
SUSY $B-L$ model with non-universal $A$-terms such that $A_N=0$ while $A_{\nu}\neq0$ 
is a promising scenario for soft leptogenesis. The proposed EDM experiments will test this 
scenario in the future.

}
\newpage
\baselineskip 24pt
\section{{\large{\bf Introduction}}}
The current measurement of the baryon-to-entropy ratio of the Universe is given by \cite{Jungman:1995bz} %
\be%
Y_B\equiv \frac{n_B}{s} = (0.87 \pm 0.02) \times 10^{-10}, %
\label{ybexp}
\ee %
where $s = 2\pi^2 g_{\star} T^3/45$ is the entropy density and
$g_\star$ is the effective number of relativistic degrees of
freedom. CP violation is an essential requirement in order to
obtain this asymmetry. It is well known that in the standard model (SM), it is not
possible to generate sufficient baryon asymmetry through the phase
of the Cabibbo-Kobayashi-Maskawa (CKM) mixing matrix, $\delta_{CKM}$ \cite{Farrar:sp}.

Supersymmetric (SUSY) extensions of the SM contain new CP-violating
sources beyond $\delta_{CKM}$, namely the Higgs bilinear term,
$\mu$, and the soft breaking terms (gaugino and squark soft
masses, bilinear and trilinear couplings). The most stringent
constraints on the SUSY phases come from continued efforts to
measure the electric dipole moments (EDM) of the neutron,
electron, and mercury atom \cite{Abel:2001vy}.

Leptogenesis \cite{yanagida}, based on a high scale seesaw mechanism, provides an
attractive scenario to explain the baryon asymmetry. However, in
this scenario, supersymmetry should be introduced to stabilize the
electroweak scale. Therefore, leptogenesis is more natural in SUSY
models. Recently, a new leptogenesis scenario, soft leptogenesis, 
has been proposed \cite{grossman,D'Ambrosio:2003wy,Grossman:2005yi}, where
sneutrino decays offer new possibilities for generating the asymmetry.

Assuming universal soft SUSY breaking terms, the relevant terms
for the soft leptogenesis in the minimal supersymmetric standard model (MSSM) extended with three right-handed neutrino superfields are given by %
\be %
{\cal L}_{\rm soft} = \frac{\tilde m_N^2}{2} \tilde{N^c}^{\dag}
\tilde{N^c}+ \frac{B_M^2}{2} \tilde{N^c} \tilde{N^c} + A_\nu
Y_{\nu} \tilde{L} \tilde{N^c} H_2 + h.c., %
\ee %
and in the case of mSUGRA, $B_M^2=B_N M_N$.
This sector has one physical CP violating phase %
\be %
\phi_\nu = {\rm arg}(A_\nu B^*_N) . %
\ee %
In this respect, a mixing between the sneutrino $\tilde{N}^c$ and
the anti-sneutrino $\tilde{N}^{c\dag}$ is an analogue to the
$B^0-\bar{B}^0$ and $K^0-\bar{K}^0$ systems. The mass difference
and the two sneutrino mass eignestates are given by %
\be%
\Delta M = \vert B_N \vert, ~~~~~~~ \Delta \Gamma = \frac{2 \vert
A_\nu \vert \Gamma}{M_N}. %
\ee %
The CP violation in the $\tilde{N}^c$-mixing, induced by the phase
$\phi_\nu$, generates lepton asymmetry in the final states of the
$\tilde{N}^c$-decay. This lepton asymmetry is converted to baryon
asymmetry through the sphaleron process \cite{spha}. The baryon to entropy
ratio for $M_N \gg A_{\nu}$ case is given by \cite{D'Ambrosio:2003wy}%
\be %
\frac{n_B}{s} \simeq - 10^{-3} \eta \left[ \frac{4 \Gamma \vert
B_N \vert }{4 \vert B_N \vert^2 + \Gamma^2} \right] \frac{\vert
A_\nu \vert}{M_N} \sin \phi_\nu, %
\label{basymmetry}
\ee %
where $\eta$ is the efficiency parameter which is suppressed for
small and large $M_N$ because of the insufficient $\tilde{N}$
production and strong washout effect.

It has been noticed \cite{D'Ambrosio:2003wy} that for 1 TeV $\ll
M_N \leq 10^8$ soft leptogenesis may give important contribution
to the baryon asymmetry only if the parameter $B_N$ is very small.
It means that, in this case, deviation from resonant condition
gives too small baryon asymmetry.  

The TeV scale right-handed neutrino is naturally obtained in
supersymmeric $B-L$ extension of the standard model (SUSY $B-L$),
which is based on the gauge group $G_{B-L}\equiv SU(3)_C \times
SU(2)_L \times U(1)_Y \times U(1)_{B-L}$. In this type of model,
the $B-L$ Higgs potential receives large radiative corrections
that induce spontaneous $B-L$ symmetry breaking at TeV scale, in
analogy to the electroweak symmetry breaking in MSSM
\cite{masiero}. This result provides further motivation for
considering the phenomenological and cosmological implications of
this model.

In this paper, we investigate the possibility of soft leptogenesis
in minimal SUSY $B-L$ model. This model has the B-term coming from  
a new A-term $A_N \tilde N \tilde N \chi_1$ and $\mu$-term $\mu' \chi_1 \chi_2$, 
where $\tilde N$ is sneutrino and $\chi_{1,2}$ are scalars which break $U(1)_{B-L}$ by their 
vacuum expectation values (VEV). 
We study the condition of the B-term for successful soft leptogenesis in 
SUSY $B-L$ model, 
which derives relation between $A_N$, $\mu'$ and $B-L$ breaking VEVs. 
We also investigate the relation between electron EDM and
soft leptogenesis, which both are generated by the same order one
phase. Electron EDM and soft leptogenesis in the MSSM 
has been studied in Ref.\cite{kashti}, 
and the result is that contribution to electron EDM from soft SUSY breaking terms is 
well suppressed and soft leptogenesis works without constraints from EDMs. 
We show that this result holds in $B-L$ model as well, because of the small Dirac neutrino Yukawa 
couplings. However, planned future experiments will test our model.   

The paper is organized as follows. In section 2 we discuss
the Minimal SUSY $B-L$ which accounts for three right-handed
neutrinos at TeV scale. In section 3 we study electron EDM from 
the CP violating phases in the (s)neutrino
sector which is responsible for soft-leptogensis. 
The analysis of soft leptogenesis in this class of models
is discussed in section 4. 
Finally we give our conclusions in section 5.

\section{{\large{\bf Supersymmetric $B-L$ extension of the SM}}}

A low scale $B-L$ symmetry breaking, based on the gauge group
$G_{B-L}\equiv SU(3)_{C}\times SU(2)_{L}\times U(1)_{Y}\times
U(1)_{B-L}$, has been recently considered
\cite{Khalil:2006yi,Abbas:2007ag,okada}. It was shown that this model
can account for the current experimental results of the light
neutrino masses and their large mixing. Therefore, it can be
considered as one of the strong candidates for minimal extensions
of the SM. Moreover, it was demonstrated that, similar to the
electroweak symmetry breaking in MSSM, the $U(1)_{B-L}$ symmetry
is radiatively broken at TeV scale in supersymmetric extension of
this class of model \cite{masiero}. Therefore, This type of models
provides a natural framework for implementing TeV seesaw
mechanism.

The part of SUSY $B-L$ superpotential, which is relevant for our
analysis, is given by
\begin{equation}
W= Y_{\nu ij}N_{i}^{c}L_{j}H_{2} - Y_{eij}E_{i}^{c}L_{j}H_{1} +
\frac{1}{2} Y_{Nij} N_{i}^{c}N_{j}^{c} \chi_1 + \mu H_{1}H_{2} +
\mu' \chi_1 \chi_2,%
\end{equation}
where $i,j=1\ldots 3$ are generation indices and the superfields $E^{c}$, $%
L=(N,E)$, $N^{c}$ contain the leptons $e_{R}^{c}$, $(\nu
_{L},e_{L})$, $\nu _{R}^{c}$, respectively. $\chi_{1,2}$ are SM gauge singlet superfields which 
break $B-L$ symmetry by their VEVs. Note that $Y_{B-L}$
for leptons and Higgs are given by $Y_{B-L}(L)=
-Y_{B-L}(E^c)=-Y_{B-L}(N^c)=-1$, $Y_{B-L}(H_1)= Y_{B-L}(H_2)=0$,
$Y_{B-L}(\chi_1)=- 2$, and $Y_{B-L}(\chi_2)=2$.

The associated soft SUSY breaking terms (assuming that SUSY
breaking scale is larger than $B-L$) are in general given by
\cite{masiero}
\begin{eqnarray}
- {\cal L}_{soft} &=&{\widetilde{m}}_{Lij}^{2}{\widetilde{L}}_{i}^{\dagger }{\widetilde{L}}_{j} + {\widetilde{m}}%
_{Eij}^{2}{\widetilde{E}}_{i}^{c* }{\widetilde{E}}_{j}^{c}  + {\widetilde{m}}_{Nij}^2 {\widetilde{N}}^{c*}_i
{\widetilde{N}}^c_j
+m_{\chi_1}^2 \left| \chi_1 \right|^2+m_{\chi_2}^2 \left| \chi_2 \right|^2
\label{soft} \\
&+& \left[Y_{\nu ij}^{A}{\widetilde{N}}_{i}^{c}
{\widetilde{L}}_{j}H_{2} - Y_{eij}^{A}{\widetilde{E}}_{i}^{c}
{\widetilde{L}}_{j}H_{1} + \frac{1}{2}
Y_{Nij}^{A}{\widetilde{N}}_i^{c} {\widetilde{N}}_j^c \chi_1 +
B\mu' \chi_1 \chi_2 \right. \nonumber\\
&+& \left.  \frac{1}{2} M_1 {\widetilde{B}}{\widetilde{B}} +
\frac{1}{2} M_2{\widetilde{W}}^a {\widetilde{W}}^a + \frac{1}{2}
M_3 {\widetilde{g}}^a {\widetilde{g}}^a + \frac{1}{2} M_{B-L}
{\widetilde{Z}_{B-L}}{\widetilde{Z}_{B-L}}+ h.c \right] \;.
\nonumber
\end{eqnarray}
Note that, due to the $B-L$
invariance, the bilinear coupling
$B_{Nij}^{2}{\widetilde{N}}_i^{c} {\widetilde{N}}_j^c $ is not 
allowed. It may be generated only after the $B-L$ symmetry breaking by 
the vacuum expectation values $\langle \chi_{1,2}\rangle=v_{1,2}'$.
In this case, $B_{N}^2$ is given by $B_N^2 = -v_1' Y_N^A +Y_N v_2' \mu'^{*}$. 

The $B-L$ minimization conditions can be used to determine the
supersymmetric parameter ${\mu^{\prime}}$ \cite{masiero}. Similar
to the electroweak breaking condition in MSSM,
one finds %
\be %
{\mu^{\prime}}^2= \frac{m_{\chi_2}^2 - m_{\chi_1}^2 \tan^2 \theta }{\tan^2\theta-1} - \frac{1}{2}M_{Z_{B-L}}^2 ,  %
\ee%
where $v'_1=v' \sin \theta,~v'_2= v' \cos \theta$ and $U(1)_{B-L}$ gauge boson mass 
$M_{Z_{B-L}}^2=8 g_{B-L}^2 {v'}^2$.

After imposing the electroweak and $B-L$ symmetry breaking
conditions, one can compute the spectrum at low energy scale and
analyze possible phenomenological consequences. Here, we present
the general expressions for the charged slepton and sneutrino mass
matrices in SUSY $B-L$. Now we adopt the super-MNS basis which,
in analogue to the super-CKM basis in the quark sector, is
defined as follows. Given the Yukawa matrices, we perform unitary
transformations of the lepton superfields $L=(N,E)$, $E^c$ and $N^c$ such
that the lepton mass matrices take diagonal forms:
\begin{eqnarray}
N_{L} &\to &V_{L}^{\nu }N_{L}\;,  \nonumber \\
E_{L,R} &\to &V_{L,R}^{e}E_{L,R}\;,
\end{eqnarray}
with $Y_{eff}^{\nu }\rightarrow (V_{L}^{\nu })^{T}Y_{eff}^{\nu }V_{L}^{\nu }=%
\mathrm{diag}(h_{\nu _{e}},h_{\nu _{\mu }},h_{\nu _{\tau }})$ and $%
Y^{e}\rightarrow (V_{R}^{e})^{\dagger }Y^{e}V_{L}^{e}=\mathrm{diag}%
(h_{e},h_{\mu },h_{\tau })$. In this basis, the leptonic charged
current interactions is given by
\begin{equation}
-\frac{g}{\sqrt{2}}\left( \overline{\ell}_{Li}\gamma ^{\mu }(V_{L}^{e\dagger
}V_{L}^{\nu })_{ij}\nu _{Lj}W_{\mu }+h.c.\right),
\end{equation}
where $g$ is the weak $SU(2)_{L}$ gauge coupling. The lepton flavour mixing
matrix is then given by
\begin{equation}
U_{MNS}=V_{L}^{e\dagger }V_{L}^{\nu }.
\end{equation}
We will assume diagonal charged lepton mass matrix, \ie, $V_L^e =
V_R^e= I$ and $V_L^\nu = U_{MNS}$.

In this class of models with TeV scale
right-(s)neutino, the low-energy sneutrino mass matrix is more
involved \cite{masiero}. It turns out that the sneutrino is
$12\times 12$ hermitian matrix. In the basis of $(\phi_{\nu_L},
\phi_{\nu_R})$ with $\phi_{\nu_L}=(\tilde{\nu}_L,
\tilde{\nu}^*_L)$ and $\phi_{\nu_R}=(\tilde{N}^c,
\tilde{N}^{c*})$, it is given by \cite{dedes}
\[
{\cal M}^{2}=\left(
\begin{array}{cc}
M^{2}_{\nu_L \nu_L} & ~~  M^{2}_{\nu_L \nu_R}
\\ \\
M^{2}_{\nu_R \nu_L}  & ~~ M^{2}_{\nu_R \nu_R}
\end{array}
\right) ,
\]
where $M^{2}_{\nu_{A} \nu_{B}}$ ($A,B \equiv L,R)$ can be written
as \cite{dedes}
\[
M_{\nu_A \nu_B}^{2}=\left(
\begin{array}{cc}
M^{2}_{A^\dag B} & ~~  M^{2^*}_{A^T B}
\\ \\
M^{2}_{A^T B} & ~~  M^{2^*}_{A^\dag B}
\end{array}
\right) ,
\]
with
\begin{eqnarray}
M^2_{\tilde{\nu}^\dag_L \tilde{\nu}_L} &=&U_{MNS}^{\dagger }\tilde{m}%
_{L}^{2}U_{MNS}+ \frac{m_{Z}^{2}}{2}\cos2\beta + v^2 \sin^2\beta
U_{MNS}^{\dagger} (Y_{\nu}^\dagger Y_\nu) U_{MNS}, \nonumber\\
M^2_{\tilde{\nu}_R^{ \dag} \tilde{\nu}_R} &=& \tilde{m}_{N}^{2} + M_N^2 
+ v^2 \sin^2 \beta (Y^*_{\nu}Y^T_{\nu}),  \nonumber \\
M^2_{\tilde{\nu}^T_L \tilde{\nu}_R} &=& -v \sin \beta~
U_{MNS}^{T} . (Y^A_{\nu})^T - v \cos\beta \mu
U_{MNS}^{T} .(Y_{\nu})^T ~,\nonumber\\
M^2_{\tilde{\nu}^{T}_R \tilde{\nu}_R} &=&M_N \mu'\cos \beta -v' \sin\theta~
Y_N^A~, \nonumber\\
M^2_{\tilde{\nu}^\dag_L \tilde{\nu}_R} &=& v \sin\beta U_{MNS}^\dag(Y_\nu)^\dag
M_N~,\nonumber\\
M^2_{\tilde{\nu}^T_L \tilde{\nu}_L} &=& 0, %
\label{sqmass1}
\end{eqnarray}
where $M_N = Y_N v' \sin \theta$.
Here few comments are in order: $i)$ In the super-MNS basis, we do not 
perform any rotation by $N_i^c$ since it is already assumed
(without loss of generality) that $M_N$ is in diagonal form. $ii)$
In the above expressions, we have kept the contribution to
Eq.(\ref{sqmass1}) proportional to the Dirac mass of the
neutrinos because in general the unitary transformation which
diagonalised $Y_{eff}^{\nu }$ doesn't necessarily diagonalise
$Y^{\nu \dagger }Y^{\nu }$. $iii)$ In general, the order of magnitude of the sneutrino mass matrix is as follows:%
\be %
{\cal M}^2 \simeq \left(\begin{array}{cc}
                {\cal O}(v^2) & {\cal O}(v v')\\
                {\cal O}(v v') & {\cal O}({v'}^2)
                \end{array}\right).
\ee %
Since $v'\sim $ TeV, the sneutrino matrix element are of the same
order and there is no a seesaw type behavior as usually found in
MSSM extended with heavy right-handed neutrinos. Therefore a
significant mixing among the left- and right- handed sneutrinos is
obtained. $iv)$ The trilinear couplings $A_{\nu}$ and $A_{N}$ are
the only SUSY sources for CP violation in the sneutrino mass
matrix (assuming $\mu$ and $\mu'$ are real). The impact of this
feature on the electron EDM and soft leptogenesis will be analyzed in
next sections. $v)$ The results of light neutrino masses can be
accommodated in this class of models with $M_N \sim {\cal O}(1)$
TeV if the neutrino Yukawa coupling $Y_{\nu}$ is of order $\lsim
10^{-6}$ \cite{Khalil:2006yi,Abbas:2007ag,masiero}, which is close
to the order of magnitude of the electron Yukawa coupling.
Therefore, if one neglects the contributions proportional to
$Y^2_\nu$, the
sneutrino mass matrix in the $(\tilde \nu_L,\tilde \nu_L^*,\tilde N^c, \tilde N^{c*})$ basis 
can be written as (where all the entries are $3\times 3 $ matrices): %
\be %
{\cal M}^2 \simeq \left(\begin{array}{cccc}
                \tilde{m}_L^2 +\frac 12 m_Z^2 \cos 2 \beta
                & 0 & v_2 Y_\nu^\dag M_N & -v_2~ U_{MNS}^\dag. (Y_\nu^A)^\dag
                \\  \\
                0  & \tilde{m}_L^2 +\frac 12 m_Z^2 \cos 2 \beta
                & -v_2~ U_{MNS}^{T}. (Y_\nu^A)^T &
                v_2 Y_\nu^T M_N\\   \\
                v_2  Y_\nu M_N & -v_2 (Y_\nu^A)^*U_{MNS}^* 
                &  \tilde{m}_N^2 + M_N^2 &- v'_1~ (Y_N^A)^*+ v'_2 Y_N \mu'\\   \\
                -v_2 (Y_\nu^A) U_{MNS}& v_2 Y_\nu^* M_N &
               - v'_1~ Y_N^A + v'_2~ Y_N \mu'^*& \tilde{m}_N^2 + M_N^2\\
                \end{array}\right).
                \label{snumass}
\ee %
It is worth noting that, in general, the trilinear coupling is not
proportional to the corresponding Yukawa coupling. Therefore, $Y_\nu^A$ in
general is not suppressed by the small $Y_\nu$. Also, the Yukawa
coupling $Y_N$ is essentially unconstrained. In this case, the
off-diagonal elements is expected to be of the same order as the
diagonal ones. Thus, a large mixing is obtained and  mass
insertion approximation is not a proper approximation in this
scenario. The above sneutrino mass matrix can
diagonalized by %
\be %
\Gamma_{\tilde{\nu}} {\cal M}^2 \Gamma_{\tilde{\nu}}^\dag = {\cal
M}^2_{diag}. %
\ee %
Hence, %
\be %
(\tilde{\nu}_{phy})_i = (\Gamma_{\tilde{\nu}})_{ij}
\tilde{\nu}_j~, ~~~~~ i,j=1,2...,12.%
\ee %

For later convenience, we mention sneutrino sector for $v_{1,2}=0$ case,   
because we will consider leptogenesis by sneutrino decay above the electroweak breaking scale.  

For $v_{1,2}=0$, off-diagonal blocks of Eq.(\ref{snumass}) vanish. 
Thus, one can easily observe that $12\times
12$ sneutrino mass matrix Eq.(\ref{snumass}) is divided to two sperate mass matrices 
for the left-handed sneutrino with no-mixing and the right-handed
sneutrino mass matrix with mixing of order $v'$. 
The $6 \times 6$ block part of the sneutrino mass matrix in the $(\tilde N^c,\tilde{N}^{c{\dag}})$ basis is 
\be %
{\cal M}_{N}^2 = \left(\begin{array}{cc}
                 \tilde{m}_N^2 + M_N^2 &B_M^{2*}~\\   \\
                B_M^2& \tilde{m}_N^2 + M_N^2\\
                \end{array}\right), 
\ee %
where $B_M^2= -v' \sin\theta Y_N A_N + v'\cos\theta Y_N \mu'=M_N (-A_N+\mu' \cot \theta)
\equiv M_N B_N$ ($\mu'$ has been assumed to be real).
Here, the mSUGRA relation $Y_N^A = Y_N A_N$ and $M_N=Y_N v' \sin \theta$ has been used. 
Now we assume the universal soft SUSY breaking terms for $\tilde m_N^2$ and $A_N$, and 
mass matrix of the heavy right-handed neutrinos $M_N$, then $B_M^2$, is already diagonal.  
One finds the mass eigenvalues of sneutrinos to be %
\be %
M^2_{\tilde{N}_{\mp i}} = M_{Ni}^2 + \tilde{m}^2_{Ni} \mp
M_{Ni} \left| B_{N} \right|,  %
\label{npmmass}
\ee %
where
\be
\left| B_{N} \right|=\sqrt{\left( -\left|A_N \right|\cos \theta_{A_N}+\mu' \cot \theta \right)^2
+\left| A_N \right|^2 \sin^2 \theta_{A_N}}, 
\label{BN}
\ee
and the mass eigenstates
\bea %
\tilde{N}_{+i}&=& \frac{1}{\sqrt{2}} \left( e^{i\phi/2} \tilde{N}^c_i + e^{-i\phi/2}{\tilde{N}_i}^{c\dag} \right)~,
\label{np}\\
\tilde{N}_{-i} &=& \frac{-i}{\sqrt{2}} \left( e^{i\phi/2}
\tilde{N}^c_i - e^{- i\phi/2} {\tilde{N}_i}^{c\dag} \right)~, %
\label{nm}
\eea %
for each generation $i=1,2, 3$. The phase of $B_M^2$, $\phi=\mbox{arg}(B_M^2)$, is given by 
\be
\phi=\tan^{-1} \left[ -\frac{\left| A_N\right| \sin \theta_{A_N}}{\left( -\left|A_N \right|\cos \theta_{A_N}+\mu' \cot \theta \right)}\right].
\label{phi}
\ee

On the other hand, the Higgs sector of this model consists of two
Higgs doublets and two Higgs singlet with no mixing \cite{masiero}. However,
after the $B-L$ symmetry breaking, one of the four degrees of
freedom contained in the two complex singlet $\chi_1$ and $\chi_2$
are swallowed in the usual way by the $Z^0_{B-L}$ to become
massive. Therefore, in addition to the usual five MSSM Higgs
bosons: neutral pseudoscalar Higgs bosons $A$, two neutral scalars
$h$ and $H$ and a charged Higgs boson $H^{\pm}$, the three new 
degrees of freedom remain physical \cite{masiero}. They form a
neutral pseudoscalar Higgs boson $A'$ and two neutral scalars $h'$
and $H'$. Their masses at tree level
are given by %
\begin{equation}
m_{A'}^2 = \mu_1^2 + \mu_2^2,
\end{equation}
\begin{equation}
m_{H',h'}^2 = \frac{1}{2}\left(m_{A'}^2 + M^2_{Z_{B-L}} \pm
\sqrt{(m_{A'}^2 + M_{Z_{B-L}}^2)^2-4{m_{A'}^{2}} M_{Z_{B-L}}^2\cos^2 2
\theta}\right).
\label{HHmass}
\end{equation}
Here
$\mu^2_{\alpha}=m^2_{\chi_\alpha} + \mu'^2$ with $\alpha=1,2$ \cite{masiero}.

The physical CP-even extra-Higgs bosons $H'_{\alpha}=(h',H')^T$ and 
CP-odd Higgs bosons $A'_{\alpha}=(G',A')^T$ are obtained from the
rotation by orthogonal matrices $O_{R(I)}$: %
\bea
H'_{\alpha}=\left( O_R\right)_{\alpha \beta}\mbox{Re} \chi_{\beta},~~
A'_{\alpha}=\left( O_I\right)_{\alpha \beta}\mbox{Im} \chi_{\beta},
\eea
where
\bea
O_R=\left(\begin{array}{cc} \cos\alpha_R &
-\sin\alpha_R \\
\sin\alpha_R & \cos\alpha_R \end{array}\right),~~
O_I=\left(\begin{array}{cc} \sin\alpha_I &
\cos\alpha_I \\
-\cos\alpha_I & \sin\alpha_I \end{array}\right), 
\eea
where the mixing angle $\alpha_R$ is given by%
\be%
\alpha_R=\frac{1}{2} \tan^{-1}\left[\tan2\theta \frac{m_{A'}^2 +
M_{Z_{B-L}}^2}{m_{A'}^2 - M_{Z_{B-L}}^2} \right]. %
\ee %

The effect of the right-sneutrino mixing on the lepton asymmetry
can be determined from the sneutrino interaction Lagrangian, which 
contains couplings of both Dirac Yukawa coupling $Y^{\nu}$ and new interaction $Y_N$. 
For $Y_{\nu}$, these are given in the basis of $(\tilde{N}_{-i},\tilde{N}_{+i})$ by %
\bea
- {\cal L}_{Y^{\nu}} &=&\frac{1}{\sqrt{2}}Y^{\nu}_{ij}e^{-i\phi/2}\tilde N_{+i}\left[ \ell^j_L \tilde H_{2L}
+\left( A_{\nu}+M_{Ni}e^{i \phi}\right)\tilde \ell^j H_2+\mu^* H^\dag_1 \tilde \ell^j \right] \nonumber \\
&+& \frac{i}{\sqrt{2}}Y^{\nu}_{ij}e^{-i\phi/2}\tilde N_{-i}\left[ \ell^j_L \tilde H_{2L}
+\left( A_{\nu}-M_{Ni}e^{i \phi}\right)\tilde \ell^j H_2+\mu^* H^\dag_1 \tilde \ell^j \right]+c.c.
\label{lag1} 
\eea

Since Eq.(\ref{lag1}) contains the complex parameter $A_{\nu}$ as well as $\phi$ defined in Eq.(\ref{phi}), 
these can generate CP violating phenomena.  
We study lepton EDMs induced by the phase of $A_{\nu}$ in the next section 
before discussing leptogenesis in section 4.
\section{EDM constraint}

The present limit of the EDM of charged leptons are
\cite{Regan:2002ta}
\bea %
d_e &<& 1.6 \times 10^{-27} e~ {\rm cm} , \\
d_{\mu} &<& 1.8 \times 10^{-19} e~ {\rm cm}.
\label{presentedm}
\eea %
This is expected to further improve in the near future to become
\cite{Lamoreaux:2001hb}
\be%
d_e < 10^{-33} e~ {\rm cm} , ~~~~~~  d_{\mu} < 10^{-25} e~ {\rm cm}.%
\label{futureedm}
\ee %
It is clear that the electron EDM provides the stringent
constraint on any new CP violating contribution. Therefore, we
will focus on the electron EDM constraint on the soft leptogenesis
phase $\theta_{A_{\nu}}$ and $\theta_{A_N}$.

The effective Hamiltonian for the EDM of the electron can be
written as \cite{Abel:2001vy}
\begin{equation}
H_{\mathrm{eff}}^{\mathrm{EDM}}=C_{1}\mathcal{O}_{1}+h.c.,
\end{equation}
where $C_{1}$ and $\mathcal{O}_{1}$ are the Wilson coefficient and
the electric dipole moment operator respectively. The operator
$\mathcal{O}_{1}$ is given by
\begin{equation}
\mathcal{O}_{1}=-\frac{i}{2}\bar{e}\sigma _{\mu \nu }\gamma _{5} e
F^{\mu \nu }.
\end{equation}
The supersymmetric contributions to the Wilson coefficient of the
electron result from the one loop penguin diagrams with neutralino
and chargino exchange. In the neutralino contribution the
selectrons are running in the loop. While the chargino diagram
involves the sneutrinos. As advocated in the previous section, the
selectron mass matrix has no dependence on the CP violating phases
of SUSY breaking terms associated with the neutrino: $A_{\nu}$ and
$A_N$ which give contribution to the soft leptogenesis. In this
respect, the neutralino contribution to the electron EDM is not
relevant for our analysis and we will focus here on the chargino
contribution only. In this case
we have %
\begin{equation}
d_{e}/e=\mathrm{Im}\left( C_{e}^{\chi ^{+}}\right) ,
\end{equation}
where $e$ is the electron electric charge. To compute the Wilson
coefficient $C_e^{\chi^+}$ and study its dependence on the phase
$\theta_{A_{\nu},A_N}$.

The chargino interactions with lepton and sneutrino are given by
\begin{eqnarray}
\mathcal{L}_{e\tilde{\nu}\chi ^{+}}= g
\sum_{k=1}^{2}\sum_{\alpha=1}^{12}\sum_{a=1}^{3}~\Big(\!\!\!\!\!&-&\!\!\!\!\!%
~V_{k1}(U_{MNS})_{ab}~\bar{e}_{L}^{a}~(\chi _{k}^{+})^{*}~\sum_{b=1}^{3}(\Gamma^\dag_{\tilde{\nu}})_{b \alpha}\tilde{\nu}%
_{phy}^{\alpha} \nonumber\\
&+& U_{k2}^{*}~[Y_{e}^{\mathrm{diag}}.U_{MNS}]_{ab}~\bar{e}%
_{R}^{a}~(\chi
_{k}^{+})^{*}~\sum_{b=1}^{3}(\Gamma^\dag_{\tilde{\nu}})_{b
\alpha}\tilde{\nu}_{phy}^{\alpha} \nonumber\\
&-& V_{k2} \left( Y_\nu\right)_{ab} \bar{e}^a_L (\chi
_{k}^{+})^{*} \sum_{b=7}^{9}(\Gamma^\dag_{\tilde{\nu}})_{b
\alpha}\tilde{\nu}_{phy}^{\alpha}~\Big).
\label{chargino}
\end{eqnarray}
The above lagrangian matches the general
form of the interaction due to the exchange of spinor $\psi_i$ and
scalar
$\phi_k$ %
\be {\cal L} = L_{ik} \bar{e}_R \psi_i \phi_k + R_{ik} \bar{e}_L
\psi_i \phi_k + h.c. %
\label{general}
\ee %
In this case, one finds that EDM is given by \cite{Ibrahim:2007fb}
\be%
d_e/e =\frac{m_i}{16\pi^2 m_k^2} {\rm Im}\left(L_{ik}
R_{ik}^*\right) \left[Q_i A\left(\frac{m_i^2}{m_k^2}\right)+ Q_k
B\left(\frac{m_i^2}{m_k^2}\right)\right],%
\ee %
where the loop functions $A$ and $B$ are given by%
\bea %
A(x) &=& \frac{1}{2(1-x)^2} \left[ 3-x + \frac{2\ln x}{1-x}
\right],\\
B(x) &=& \frac{1}{2(1-x)^2} \left[ 1 + x + \frac{2 x \ln x}{1-x}
\right]. %
\eea

\begin{figure}[t]
\begin{center}
\epsfig{file=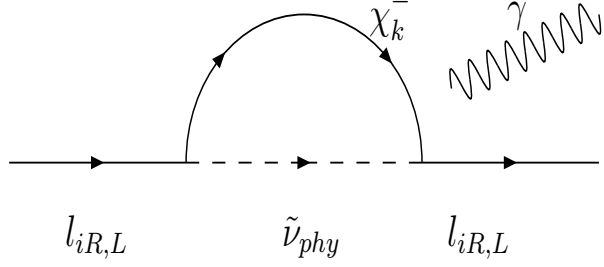,width=8cm,height=3.5cm}%
\caption{The chargino contributions to the charged lepton EDM due
to chargino and sneutrino exchange.}
\label{edmfig}
\end{center}
\end{figure}

The SUSY contribution to $d_e$ due the exchange of $\chi_i$ and
$\tilde{\nu}_k$ is shown in Fig.\ref{edmfig}.
%
%
From Eqs.(\ref{chargino},\ref{general}), one can identify $L$ and
$R$ coefficients as %
\bea %
L_{k\alpha} & = & h_e U_{k2}^{*}~ (U_{MNS})_{1b}
(\Gamma^\dag_{\tilde{\nu}})_{b \alpha}\label{L},\\
R_{k\alpha}& = & - g V_{k1}(U_{MNS})_{1b}~(\Gamma^\dag_{\tilde{\nu}})_{b\alpha} 
-V_{k2}(Y_{\nu})_{1b} (\Gamma^\dag_{\tilde{\nu}})_{(b+6)\alpha}. %
\label{R}
\eea
Therefore, the electron EDM is given by%
\be%
d_e/e =\sum_{k=1}^2 \sum_{\alpha=1}^{12} Q_{\chi_k}
\frac{m_{\chi_k}}{16\pi^2 m_{\tilde{\nu}_{\alpha}^2}} {\rm
Im}\left(L_{k\alpha}
R_{k\alpha}^*\right)  A\left(\frac{m_{\chi_k^2}}{m_{\tilde{\nu}_{\alpha}}^2}\right).%
\label{de}
\ee 
Here we assume that phase in ${\cal M}^2$ of Eq.(\ref{snumass}), that is, phase in the diagonalization matrix 
$\Gamma_{\tilde \nu}$, say $\sin \phi_{\Gamma}$, is the only origin of the contribution to the EDM. 
However, from Eqs.(\ref{L}), (\ref{R}) and (\ref{de}), the combination 
$\mbox{Im}(LR^*)$ vanishes for the first term of $R_{k\alpha}$ because 
$\Gamma_{\tilde \nu}$ dependence of $R_{k \alpha}$ is the same as that of $L_{k \alpha}$.  
The second term of $R_{k \alpha}$ gives non-zero contribution to electron EDM, but 
it is small because of suppression by Dirac neutrino Yukawa coupling $Y_{\nu}\sim10^{-6}$. 
To estimate its magnitude, we assume $(\Gamma_{\tilde \nu}^{\dag})_{b\alpha}(\Gamma_{\tilde \nu})_{\alpha (b+6)} \sim e^{i \phi_{\Gamma}}$, $U_{MNS},U,V \sim 1$ and $m_{\chi}=m_{\tilde \nu}=100 {\mbox{GeV}}(1\mbox{TeV})$, the eEDM is   
\be
d_e/e \sim  10^{-31(32)} \sin \phi_{\Gamma}~\mbox{ cm}, 
\ee
which is four (five) orders of magnitude smaller than the present experimental limit. 
Therefore eEDM does not constrain the phase $\sin \phi_{\Gamma}$ today, 
that is, $\theta_{A_{\nu}(A_N)}$, 
which is essential for soft leptogenesis as we will show in the next section. 
However the planned eEDM experiment Eq.(\ref{futureedm}) will give non-trivial tests of our 
scenario in the future.  
\section{Soft leptogenesis in SUSY $B-L$ model}
In SUSY $B-L$ extension of the SM, the neutrino Yukawa coupling
$Y_{\nu}$ is very tiny, therefore the standard thermal leptogenesis
can not account for baryon asymmetry in the Universe unless a
highly degenerate right-handed neutrino masses are assumed \cite{resonant,Abbas:2007ag}. 

Recently, a new source of lepton asymmetry, due to the induced
mixing between sneutrino-antisneutrino, has been analyzed
\cite{grossman,D'Ambrosio:2003wy,Grossman:2005yi}. In this framework, the
CP asymmetry in decay of the heavy sneutrino $\tilde N_-\equiv \tilde N_{-1}$ defined in Eq.(\ref{np}) and (\ref{nm}) is given by%
\bea %
\epsilon_- &=& \frac{\sum_f \left[\Gamma(\tilde{N}_{-}\rightarrow
f) - \Gamma(\tilde{N}_{-} \rightarrow \bar{f})\right]}{\sum_f
\left[\Gamma(\tilde{N}_{-}\rightarrow f) + \Gamma(\tilde{N}_{-}
\rightarrow \bar{f})\right]}\nonumber \\%
&=&\frac{2(M_-^2-M_+^2)\Pi_{+-}\sum_f\mbox{Im}(f_-^*f_+)c_f}
{\sum_f \left[ |f_+|^2(M_-^2-M_+^2)^2+|f_- \Pi_{++}-f_+ \Pi_{+-}|^2\right]}, 
\label{ep}
\eea %
where $f$ is a final state with lepton number equal to $1$ and
$\bar{f}$ is its conjugate. In the second equation, $f_{\pm}$ are tree-level decay amplitudes and 
$\Pi_{\pm \pm,\pm \mp}$ are the absorptive part of 
two point functions, which are given below. 
The factor $c_f (f=B,F$ for bosonic and fermionic final state) is introduced to 
parametrize the phase space of the bosonic and fermionic 
final states. Asymmetry by $\tilde N_+$ is obtained by exchanging $+ \leftrightarrow-$ in Eq.(\ref{ep}).
 The effect of the
$\tilde{N}^c-\tilde{N}^{c\dag}$ mixing on the lepton asymmetry
$\epsilon$, which is assumed to be dominated the direct CP
violation in this decay, can be determined by computing the
sneutrino mass eigenstates.

\subsection{MSSM+$N_1$ case}
First we briefly mention the MSSM+heavy right handed (s)neutrino $N_1 (\tilde N_1)$ case with 
$M_N \gg \mbox{TeV}$ \cite{D'Ambrosio:2003wy}. In this model, 
CP asymmetry from heavy sneutrino $\tilde N_{\pm}$ decay into $\ell \tilde H$ and $\tilde \ell H$ is given by
\bea
\epsilon =\frac{4 \Gamma B_N}{4 B_N^2 +\Gamma^2}\frac{\mbox{Im}A_{\nu}}{M_N}\Delta_{BF},
\eea
where the total decay rate $\Gamma$ is 
\bea
\Gamma=\frac{(Y_{\nu}Y_{\nu}^{\dag})_{11}}{4 \pi}M_N, 
\label{gammaMSSM}
\eea
and $\Delta_{BF}\equiv (c_B-c_F)/(c_B+c_F)$. This gives the largest value of $\epsilon$ at 
$\Gamma=2B_N$, which is the resonance condition, and therefore 
\bea
M_N=\left( \frac{10^{-3}\mbox{eV}}{m_{\nu}}\right)^{1/2}
\left( \frac{B_N}{100 \mbox{GeV}}\right)^{1/2}~10^{10}~\mbox{GeV}
<\frac{\mbox{Im}A_{\nu}}{1\mbox{TeV}}~10^{8-9}\mbox{GeV}. 
\eea 
The equality comes from the resonance condition, and the inequality from 
that $\epsilon$ is large enough to obtain observed baryon asymmetry. 
One can see that this requires small b-term: $B_N \sim{\cal O}(10)\mbox{GeV}$.

\subsection{$B-L$ case}
Next we discuss our $U(1)_{B-L}$ model. 
It is worth mentioning that the
leptogenessis process takes place at large scale, around $B-L$ breaking 
scale ($v'$). At this scale, the electroweak symmetry is still an exact
symmetry, \ie, $v=0$. Thus, as given in section 2, one can easily observe that $12\times
12$ sneutrino mass matrix Eq.(\ref{snumass}) is divided to two sperate mass matrix
for left-handed sneutrino with no-mixing and right-handed
sneutrino mass matrix with mixing of order $v'$. 
We will focus on
the lightest right-sneutrino $\tilde{N}_{1}$, and therefore its 
mass eigenstate $\tilde N_{\pm}$, since the lepton
asymmetry is usually dominated by the decay of the lightest one. 

We consider soft leptogenesis in this model. 
From Eq.(\ref{lag1}), CP asymmetry is generated by decay processes of 
the lightest heavy sneutrino $\tilde N_{\pm}$ into $\tilde \ell+H$ and $\ell+\tilde H$. 
Moreover, if heavy neutrinos $N_i$ and new particles $H'_{\alpha},A'_{\alpha}$ and $\chi_{1,2}$    
are lighter than $\tilde N_{\pm}$,  there are other decay modes: 
$\tilde N_{\pm} \to N_i+\chi_{1}$ and $\tilde N_{+} \to \tilde N_- + (H',A')$. 
However, if the new particles are heavier than $\tilde N_{\pm}$, these decay modes are kinematically 
forbidden and can not contribute to CP asymmetry. 
In the following analysis, we neglect these new processes by assuming 
these new particles are heavier than $\tilde N_{\pm}$. 
 
In this case, the total decay width of $\tilde{N}_-$ is given by %
\be %
\Gamma_- = 2 \Gamma(\tilde{N}_- \to \ell+\tilde{H_2}) 
+ 2 \Gamma(\tilde{N}_-\to \tilde{\ell}+H_2)+
2 \Gamma (\tilde N_- \to \tilde \ell+H_1),%
\ee%
where the factor $2$ comes from decay into anti-particles. 
The tree level contribution to the decay of $\tilde{N}_-$ to
Higgs doublet and charged lepton doublet is given by %
\begin{eqnarray} %
\Gamma(\tilde{N}_- \to \ell+ \tilde{H}_2) &=& 
\frac{1}{2 M_-}\sum_i\left| f_{-}(\ell^i \tilde H_2)\right|^2
I_2(M_-; m_{\ell^i},m_{\tilde H_2}), 
\label{gammaBL1}\\ %
\Gamma(\tilde{N}_- \to \tilde{\ell}+ H_2) &=&
\frac{1}{2M_-}\sum_i \left| f_-(\tilde \ell^iH_2)\right|^2
I_2(M_-; m_{\tilde \ell^i},m_{H_2}), 
\label{gammaBL2} \\
\Gamma(\tilde N_- \to \tilde \ell+H_1)&=&
\frac{1}{2M_-}\sum_i \left| f_-(\tilde \ell^iH_1)\right|^2
I_2(M_-; m_{\tilde \ell^i},m_{H_1}), 
\label{gammaBL3}
\end{eqnarray} %
where phase space integral $I_2$ is 
\bea
I_2(x;y,z)=\frac{1}{8\pi x^2}\sqrt{\left[ x^2-(y-z)^2\right]\left[ x^2-(y+z)^2\right]}, 
\eea
and $- \to +$ for $\tilde N_{+}$ decay. 
Tree-level amplitudes $f_{\pm}$ are defined as 
\bea
f_{\pm}(\ell^i \tilde H_2)&=&-i Y^{\nu}_{1i}e^{i \phi/2}\sqrt{M_{\pm}^2-m_{\tilde H_2}^2-m_{\ell^i}},
\label{fpm1}\\
f_{\pm}(\tilde \ell^iH_2)&=&-i Y^{\nu}_{1i}e^{i \phi/2}\left( A_{\nu}^* \pm M_N e^{-i \phi}\right),\\
f_{\pm}(\tilde \ell^i H_1)&=&-iY^{\nu}_{1i} e^{i \phi/2} \mu. 
\label{fpm6}
\eea

In order to satisfy the out of equilibrium condition, we should
have $\Gamma < H(M_{-})$, where $H(M_{-})$ is the Hubble parameter
at temperature $T=M_-$, namely %
\be %
H(M_-) \simeq \frac{g_*^{1/2} M_-^2}{M_{pl}}~. %
\ee %
If $M_-$ is of order $10^3$ GeV (\ie, it is dominated by soft scalar
mass $\tilde{m}_N$, which can be of that order), then $H(M_-)
\simeq 10^{-12}$ GeV, hence $\Gamma \lsim 10^{-12}$ GeV is required. 
However for $M_N \simeq 10^3$ GeV and $Y_{\nu} \sim 10^{-6}$, 
the ratio $\Gamma/H \sim 10$, hence efficiency factor can be estimated to be $\eta \sim 0.1$. 
Then, baryon asymmetry is given by 
\be
Y_B=-8.6 \times 10^{-4}\epsilon ~\eta.
\ee

\begin{figure}[h,t]
\begin{center}
\epsfig{file=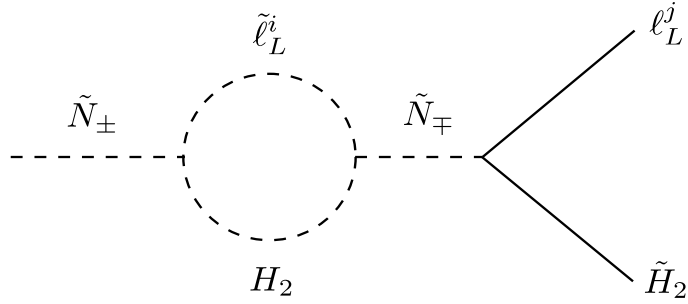,width=9cm,height=4cm,angle=0}
\end{center}
\caption{The contributions to the decay $\tilde{N}_- \to
\ell_L^j \tilde{H_2}$. The bosonic final states $\tilde \ell_L^j H_2$ and $\tilde \ell_L^j H_1$ are also possible.}
\label{fig:B-Ldiagram}
\end{figure}

The CP asymmetry $\epsilon$ is generated by interference between the tree and 
the following self-energy diagram (Fig.\ref{fig:B-Ldiagram}) with 
a bosonic loop. Diagrams of fermionic loops vanishes. 
 
%


The two-point functions $\Pi$ obtained from the diagram Fig.\ref{fig:B-Ldiagram} are 
given by
\bea
\Pi_{++}&=&M_+ \Gamma_+,\\
\Pi_{--}&=&M_- \Gamma_-,\\
\Pi_{+-}&=&2 |Y^{\nu}_{1i}|^2 |A_{\nu}|^2 M_N \sin(\phi-\theta_{A_{\nu}})
I_2(M_+; m_{\tilde \ell^i},m_{H_2}),\\
\Pi_{-+}&=&2 |Y^{\nu}_{1i}|^2 |A_{\nu}|^2 M_N \sin(\phi-\theta_{A_{\nu}})
I_2(M_-; m_{\tilde \ell^i},m_{H_2}).
\eea

The masses of sneutrinos $\tilde N_{\pm}$ must be strongly degenerate in order to obtain 
enough baryon asymmetry.
Since the mass difference of $\tilde N_{\pm}$ is $M_+^2-M_-^2=M_N \left| B_N\right|$ from 
Eq.(\ref{npmmass}), the resonance condition becomes 
\be
M_N \left| B_N\right|\sim \Pi_{\pm \pm} \sim 10^{-8}\mbox{GeV}^2. 
\label{reso}
\ee
This implies that $|B_N|$ has to be extremely small, $\sim 10^{-11}$ GeV, in order to 
satisfy the resonance condition. This means that more strict degeneracy between the heavy 
sneutrino masses is required in the $B-L$ model comparing with the MSSM+$N_1$ case.   
Although the resonance condition $\Gamma \sim B_N$ 
itself is the same, the value of $\Gamma$ is different. One can see that 
$\Gamma_{MSSM+N}\gg\Gamma_{B-L}$ 
from Eqs.(\ref{gammaMSSM}) and (\ref{gammaBL1}),(\ref{gammaBL2}),(\ref{gammaBL3}) 
because $M_N \sim 1$ TeV in the $B-L$ model. 

\begin{figure}[t]
\unitlength=1mm
\begin{center}
\begin{picture}(90,50)
\includegraphics[width=8cm]{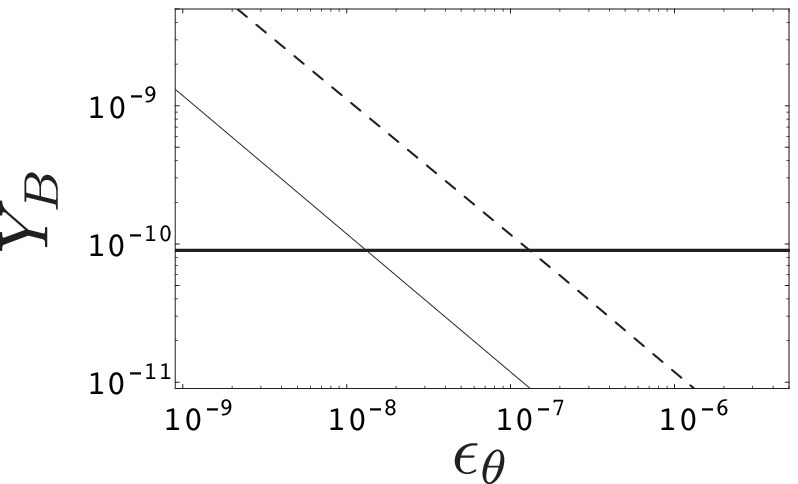}
\end{picture}
\hspace{-0.5cm}
\begin{picture}(80,50)
\includegraphics[width=8cm]{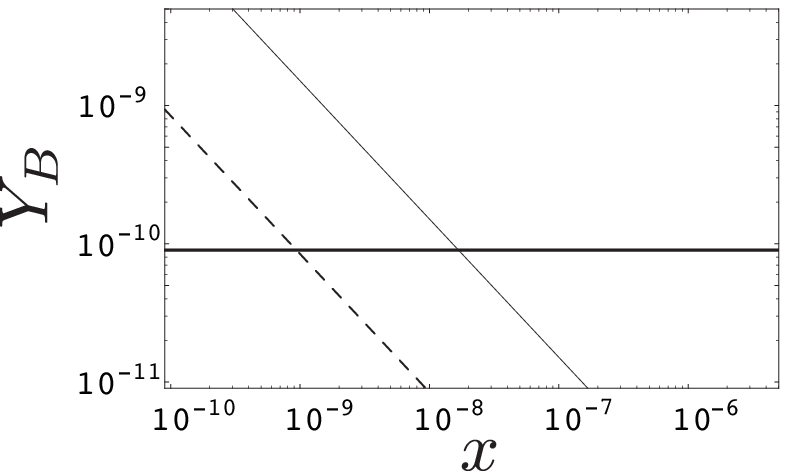}
\end{picture}
\end{center}
\caption{Baryon asymmetry as a function of new parameters 
$\epsilon_{\theta}$ (left) and $x$ (right) defined in Eqs.(\ref{epth}) and (\ref{xx}). 
In both figures, $M_N=1~{\rm TeV},~\mu'=500$ GeV and $\theta_{A_\nu}=\pi/2$. 
The horizontal line is the experimental value Eq.(\ref{ybexp}).
}
\label{y}
\end{figure}

From the definition of $B_N$, Eq.(\ref{BN}), parameters $A_N,\mu'$ and $\cot \theta=v_2'/v_1'$ have to 
be related to each other to satisfy the resonance condition Eq.(\ref{reso}); 
$|A_N|\cos \theta_{A_N} \simeq \mu' \cot \theta$ for the first term of $B_N$, and 
$\theta_{A_N} \ll 1$ and/or $|A_N| \ll {\cal O}(\rm{TeV})$ for the second term. 
While there are two possibilities for the second term, the condition for the first term is the same and 
this gives constraint on $\cot \theta$. 
\begin{enumerate}
\item $\sin \theta_{A_N}\ll 1,~|A_N|\sim {\cal O}(\rm{TeV})$ case; \\
In this case, we obtain resonance condition for a new parameter $\epsilon_{\theta}$ which  
parametrizes the deviation of $\cot \theta$ from $|A_N| \cos \theta_{A_N}/\mu'$ defined as 
\be
\epsilon_{\theta}\equiv1-\frac{|A_N| \cos \theta_{A_N}}{\mu' \cot\theta} \ll1. 
\label{epth}
\ee
The left panel of Fig.\ref{y} shows the baryon asymmetry as a function of $\epsilon_{\theta}$ for 
$|A_N|=10^3$ GeV (solid) and $10^2$ GeV (dashed), assuming $\theta_{A_N}=0$ 
and $|A_{\nu}|=10^3$ GeV. 
\item $|A_N|\ll {\cal O}(\rm{TeV})$ case;\\
For the small $|A_N|$ case, $\theta_{A_N}$ dependence is not important and 
$\theta \simeq \pi/2$ is required when $\mu'$ is large ($\gsim 10^2$ GeV).  
The resonance condition for a new parameter $x$ which parametrizes the deviation 
of $\theta$ from $\pi/2$ is  
\be
x \equiv \frac{\pi}{2}-\theta \ll 1. 
\label{xx}
\ee  
The right panel of Fig.\ref{y} shows the baryon asymmetry as a function of $x$ for 
$|A_{\nu}|=10^3$ GeV (solid) and $10^2$ GeV (dashed), assuming 
$|A_N|=0$.  
\end{enumerate}

From these figures one can see that 
enough baryon asymmetry is generated when 
$\epsilon_{\theta}\lsim10^{-7}$ or $x \lsim 10^{-8}$. Namely, 
the resonance condition for SUSY $B-L$ model,
\be
\frac{v'_2}{v'_1}=\frac{|A_N|}{\mu'},
\label{tune}
\ee
must be accurately satisfied for both $\theta_{A_N}\ll1$ and/or $|A_N|\ll {\cal O}({\rm TeV})$ cases. 
It is hard to realize the condition Eq.(\ref{tune}) for general $A_N$ and $\mu'$. 
However, there are several SUSY breaking scenarios yielding non-universal A-terms, 
and it is quite plausible to find a SUSY models with $|A_N|=0$ and $A_{\nu}\neq 0$ \cite{tatsuo}. 
In fact, in the modulus-dominated SUSY breaking case, A-terms are obtained as 
\begin{eqnarray}
A_{ijk}=-\sqrt{3}m_{3/2}(3+n_i+n_j+n_k),
\end{eqnarray}
where $m_{3/2}$ is the gravitino mass and $n_i,n_j$ and $n_k$ are modular weight of the 
fields to couple. Since the A-terms depend on the fields, these are in general non-universal. 
If we assign modular weight $n_i=-1$ for the heavy sneutrinos $\tilde N$ and singlet scalar $\chi_1$, 
then $A_N=0$. On the other hand, $n=-2$ for $\tilde L$ and $n=-3$ for $H_2$ give 
$A_{\nu}=3 \sqrt{3}m_{3/2}e^{-i\alpha}$, where $\alpha$ is the corresponding CP violating phase. 
The detailed phenomenological implications of this class of models have been analyzed in 
Ref.\cite{tatsuo}. 
This type of model is a promising scenario for implementing soft leptogenesis.
\section{Conclusions}
We have studied electron EDM and soft leptogenesis induced by 
trilinear soft SUSY breaking terms in a $B-L$ extension of supersymmetric standard model. 
The $B-L$ symmetry is broken by VEVs of extra Higgs bosons at TeV scale and 
neutrino Dirac Yukawa coupling is of order $10^{-6}$ in this TeV scale seesaw model. 
Because of the smallness of Dirac neutrino Yukawa couplings, 
the electron EDM from higgsino loop is enough suppressed, and not impose 
constraint on the CP violating phases of the trilinear terms $A_{\nu,N}$. The soft leptogenesis is also generated by the same trilinear terms.  
Since the decay rate $\Gamma$ of heavy sneutrino depends on the seesaw scale, 
the resonance condition $\Gamma \sim B_N$ requires 
small bilinear term $|B_N|\sim10^{-11}$ GeV for TeV scale seesaw model.   
This resonance condition leads to relation between $\cot \theta,~\mu'$ and $A_N$.
SUSY model of non-universal A-terms such that $A_N=0$ while $A_{\nu}\neq0$ with 
large $\tan \theta$ is a promising scenario for successful soft leptogenesis. 
We have shown that this can be naturally realized in models with non-universal soft SUSY breaking terms, as in, for example,  orbifold string models with an appropriate assignment of modular weight for each field. 
We have emphasized that this scenario has also testable implications at future collider. 
The future experiments of electron EDM will provide a serious test for the soft-letpgenesis in this class of SUSY models.

\vspace{0.5cm}
\noindent
{\large \bf Acknowledgments }\\
The authors would like to thank J. Ellis for useful discussions. 
This work is supported by the ESF grant No. 8090 (Y.K. and M.R.). 
The work of S.K. is supported in part by ICTP project 30 and the Egyptian Academy of Scientific Research and Technology.   


\bibliographystyle{unsrt}

\begin{thebibliography}{99}
\bibitem{Jungman:1995bz}
G.~Jungman, M.~Kamionkowski, A.~Kosowsky and D.~N.~Spergel,
Phys.\ Rev.\ D {\bf 54}, 1332 (1996) [arXiv:astro-ph/9512139];
M.~Zaldarriaga, D.~N.~Spergel and U.~Seljak,
Astrophys.\ J.\  {\bf 488}, 1 (1997) [arXiv:astro-ph/9702157];
P.~de Bernardis {\it et al.},
Astrophys.\ J.\  {\bf 564}, 559 (2002) [arXiv:astro-ph/0105296];
C.~Pryke, N.~W.~Halverson, E.~M.~Leitch, J.~Kovac,
J.~E.~Carlstrom, W.~L.~Holzapfel and M.~Dragovan,
Astrophys.\ J.\  {\bf 568}, 46 (2002) [arXiv:astro-ph/0104490]; 
%
WMAP Collaboration (E. Komatsu {\it et al}.), 
Astrophys. J. Suppl. {\bf 180}, 330 (2009) 
[arXiv:0803.0547 [astro-ph]];  
WMAP Collaboration (J. Dunkley {\it et al}.), 
 Astrophys. J. Suppl. {\bf 180}, 306 (2009) 
 [arXiv:0803.0586 [astro-ph]]. 
\bibitem{Farrar:sp}
G.~R.~Farrar and M.~E.~Shaposhnikov,
Phys.\ Rev.\ Lett.\  {\bf 70}, 2833 (1993) [Erratum-ibid.\  {\bf
71}, 210 (1993)] [arXiv:hep-ph/9305274];
M.~B.~Gavela, M.~Lozano, J.~Orloff and O.~P{\`e}ne,
Nucl.\ Phys.\ B {\bf 430}, 345 (1994) [arXiv:hep-ph/9406288];
M.~B.~Gavela, P.~Hern{\'a}ndez, J.~Orloff, O.~P{\`e}ne and
C.~Quimbay,
Nucl.\ Phys.\ B {\bf 430}, 382 (1994) [arXiv:hep-ph/9406289];
P.~Huet and E.~Sather,
Phys.\ Rev.\ D {\bf 51}, 379 (1995) [arXiv:hep-ph/9404302].
%
\bibitem{Abel:2001vy}
  S.~Abel, S.~Khalil and O.~Lebedev,
  Nucl.\ Phys.\  B {\bf 606}, 151 (2001)
  [arXiv:hep-ph/0103320].
%
\bibitem{yanagida}
M. Fukugita and T. Yanagida, 
Phys. Lett. {\bf B174}, 45 (1986). 

\bibitem{grossman}
Y.~Grossman, T.~Kashti, Y.~Nir and E.~Roulet, 
Phys. Rev. Lett. {\bf 91}, 251801 (2003) [arXiv:hep- 
ph/0307081]. 

\bibitem{D'Ambrosio:2003wy}
  G.~D'Ambrosio, G.~F.~Giudice and M.~Raidal,
  Phys.\ Lett.\  B {\bf 575}, 75 (2003)
  [arXiv:hep-ph/0308031].
  %
  \bibitem{Grossman:2005yi}
  E.~J.~Chun,
  Phys.\ Rev.\  D {\bf 69}, 117303 (2004)
  [arXiv:hep-ph/0404029];
 Y.~Grossman, T.~Kashti, Y.~Nir and E.~Roulet,
  JHEP {\bf 0411}, 080 (2004)
  [arXiv:hep-ph/0407063];
  Y.~Grossman, R.~Kitano and H.~Murayama,
  JHEP {\bf 0506}, 058 (2005)
  [arXiv:hep-ph/0504160].
%

\bibitem{spha}
V. A. Kuzmin, V. A. Rubakov and M. E. Shaposhnikov, 
Phys. Lett. {\bf 155}, 36 (1985). 
\bibitem{masiero}
S. Khalil and A. Masiero, 
Phys. Lett. {\bf B665}, 374 (2008) 
[arXiv:0710.3525 [hep-ph]].
%
\bibitem{kashti}
T. Kashti, 
Phys. Rev. {\bf D71}, 013008 (2005) [arXiv:hep-ph/0410319]. 
\bibitem{Khalil:2006yi}
  S.~Khalil,
  J. Phys. {\bf G35}, 055001 (2008) 
  [arXiv:hep-ph/0611205];
  W.~Emam and S.~Khalil,
  Eur. Phys. J. {\bf C522}, 625 (2007) 
  [arXiv:0704.1395 [hep-ph]]. 

\bibitem{Abbas:2007ag}
  M.~Abbas and S.~Khalil,
  JHEP {\bf 0804}, 056 (2008) 
 [arXiv:0707.0841 [hep-ph]].
\bibitem{okada}
S. Khalil and H. Okada, 
arXiv:0810.4573 [hep-ph]. 
\bibitem{dedes}
  A.~Dedes, H.~E.~Haber and J.~Rosiek,
  JHEP {\bf 0711}, 059 (2007)  
 [arXiv:0707.3718 [hep-ph]].
%
\bibitem{Regan:2002ta}
  B.~C.~Regan, E.~D.~Commins, C.~J.~Schmidt and D.~DeMille,
  Phys.\ Rev.\ Lett.\  {\bf 88} (2002) 071805; 
Muon $(g-2)$ Collaboration (G. W. Bennett {\it et al}.), 
arXiv:0811.1207 [hep-ex].
\bibitem{Lamoreaux:2001hb}
  S.~K.~Lamoreaux,
  Phys. Rev. {\bf A66}, 022109 (2002) 
  [arXiv:nucl-ex/0109014]; 
  A. Adelmann, K. Kirch, 
  arXiv:hep-ex/0606034.
\bibitem{Ibrahim:2007fb}
  T.~Ibrahim and P.~Nath,
  Rev. Mod. Phys. {\bf 80}, 577 (2008) 
 [arXiv:0705.2008 [hep-ph]].
\bibitem{resonant}
M. Flanz, E. A. Paschos and U. Sarkar, 
Phys. Lett. {\bf B345}, 248 (1995), {\bf E382}, 447 (1996); 
M. Flanz, E. A. Paschos, U. Sarkar and J. Weiss, Phys. Lett. {\bf B389}, 693 (1996); 
L. Covi, E. Roulet and F. Vissani, 
Phys. Lett. {\bf B384}, 169 (1996); 
A. Pilaftsis, 
Phys. Rev. {\bf D56}, 5431 (1997).








\bibitem{tatsuo}
  Y.~Kawamura, S.~Khalil and T.~Kobayashi,
  Nucl.\ Phys.\  B {\bf 502}, 37 (1997)
  [arXiv:hep-ph/9703239];  S.~Khalil and T.~Kobayashi,
  Nucl.\ Phys.\  B {\bf 526}, 99 (1998)
  [arXiv:hep-ph/9706479].


\end{thebibliography}

\end{document}